\documentclass[a4paper,10pt]{article}

\input epsf.tex

\newcommand{\be}{\begin{equation}}
\newcommand{\ee}{\end{equation}}
\newcommand{\ba}{\begin{eqnarray}}
\newcommand{\ea}{\end{eqnarray}}
\newcommand{\ban}{\begin{eqnarray*}}
\newcommand{\ean}{\end{eqnarray*}}

\begin{document}

\title{\Large \sc{The speed of quantum information and the preferred frame: 
analysis of experimental data}}

{\normalsize{\author{
Valerio Scarani\thanks{corresponding author; e-mail:
valerio.scarani@physics.unige.ch}, Wolfgang Tittel, Hugo Zbinden, Nicolas Gisin \\
Group of Applied Physics, University of Geneva\\
20, rue de l'Ecole-de-M\'edecine, CH-1211 Geneva, Switzerland\\
 }}}
\maketitle

\begin{abstract}
The results of EPR experiments performed in Geneva are analyzed in
the frame of the cosmic microwave background radiation, generally
considered as a good candidate for playing the role of preferred
frame. We set a lower bound for the speed of quantum information
in this frame at $1.5\times 10^4$c.           
\end{abstract}

\section{Introduction}

The tension between quantum mechanics (QM) and relativity
manifests itself in two classes of theoretical problems. The
first class of problems can be labelled "the search for a
covariant description of the measurement process". Possibly the
best-known example is the impossibility of a causal description
of the collapse in an EPR experiment that would be valid in all
frames; but there are many other examples, even for one-particle
measurements, as widely discussed by Aharonov and Albert
\cite{aha,bre}. The second class of problems is linked with some
structural problems of quantum relativistic theories, like the
definition of a position operator that fulfills "basic"
requirements \cite{reuse,caban,bac}.

In this paper, we are concerned with the first of these classes.
Actually, from an "orthodox" standpoint, these problems have been
solved \cite{aha,per,omn}: two observers, each using quantum
prescriptions, predict the same final probabilities
--- recall that in its orthodox interpretation, QM deals only with
probabilities, while the state vector and its evolution are not
endowed with reality \cite{peres}. But several physicists are not
satisfied with this solution, for different reasons \cite{mau}.
For us, the tension between the notion of event that appears in
relativity, and the reversible evolution of the quantum state, may
be a guide for new physics.

The introduction of a preferred frame (PF) is a way out of the
first class of problems, that would allow a realistic (obviously
non-local) description of the quantum measurement \cite{har}.
Moreover, it seems that the PF would also be a way out of the
second class of problems \cite{reuse,caban}. To our knowledge,
the introduction of a PF is still an intellectual tool (or
trick): no experiments are planned or even proposed that aim to
falsify this hypothesis \cite{note0}. An experiment can be easily
conceived that falsifies a {\em joint hypothesis}: suppose that
there is a PF {\em and} that in the PF the speed of quantum
information is finite, though superluminal \cite{ebe}. Then, if
quasi-simultaneity in the PF is achieved in an EPR experiment,
the EPR correlations should disappear.

In this paper we throw some light on this question by analysing
long-distance EPR experiments performed between two telecom
stations (Bellevue and Bernex) separated by 10.6 km \cite{roue}.
The main idea is that, having observed standard EPR correlations,
we are able to set a lower bound for the speed of quantum
information \cite{note1} in any given frame. The structure of the
paper is as follows. In Section 2, we define the speed of quantum
information, and give its transformation law under a Lorentz
boost. In Section 3, we introduce a good candidate for the PF,
namely, the frame of the cosmic microwave background radiation
(CMB), and give its speed with respect to the rest frame of our
laboratory (G-frame, where G stands for Geneva). The results of
these two sections are combined in Section 4 with experimental
data, leading to the announced lower bound for the speed of
quantum information in the CMB-frame. Section 5 is a conclusion.

\section{The speed of quantum information}

\begin{figure}
\begin{center}
\epsfxsize=15cm \epsfbox{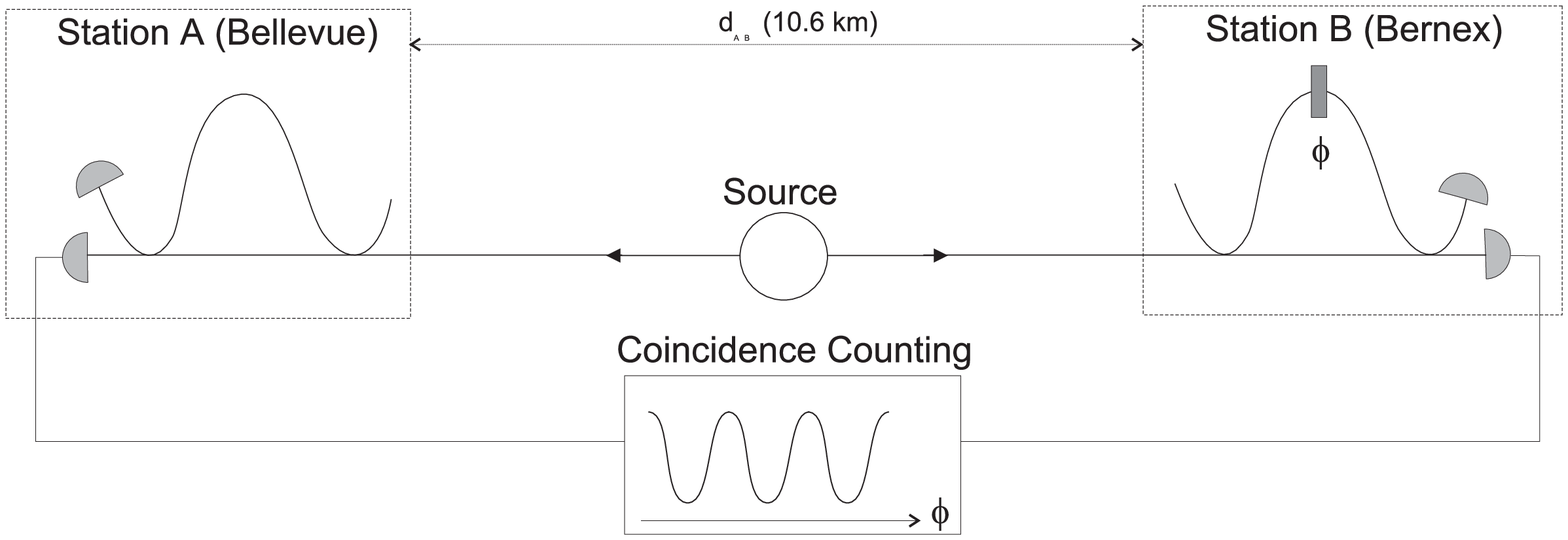}
\caption{Schematic of the experiment consisting on a photon pair
source and two analyzers separated by more than 10 km.}
\label{figsetup}
\end{center}
\end{figure}

In an optical EPR experiment (figure \ref{figsetup}), two photons
are produced in an entangled state and sent to two analyzing
stations A and B. The quantum entanglement manifests itself by the
interference fringes that are observed in the coincidence counts
of the detectors in A and B. These interferences are predicted by
QM; still, many physicists are not at ease with correlations that
arise between two space-like separated events. The correlation are
sometimes considered as due to a "superluminal influence" that the
first particle to reach its detector sends to the second one. In
this work, we call "speed of quantum information" $\vec{v}_{QI}$
the superluminal speed at which this "influence" should propagate
from one station to the other one. Of course, if two events are
space-like separated, there is always a frame in which the two
events are simultaneous ($v_{QI}=\infty$), and a family of frames
in which the ordering of the arrivals is inversed with respect to
the laboratory frame. Therefore the supposed "superluminal
influence" is a real physical process only in a preferred-frame
theory, or in a theory in which the meaningful frames are not
arbitrary \cite{sua}. However, the operational definition of
$\vec{v}_{QI}$ involves events (detections), that can be
parametrized by using the standard relativistic formalism; in
other words, a speed of quantum information can be defined
formally in any frame. Its definition goes as follows: since
correlations were observed, the quantum information must have
travelled the distance between the two detectors in the time
interval between the two detections. Let's define the $x$ axis as
the axis linking the detector in $A$ and the detector in $B$,
oriented from $A$ to $B$. Therefore, if the event "detection at A"
is parametrized in a given frame by $(x_A,t_A)$, and similarly for
the event "detection in B", the speed of quantum information must
have the same direction as and have a higher value than \be
v_{QI,min}\,=\,\frac{x_A-x_B}{t_A-t_B}\,.\ee Let $\vec{v}$ be the
speed of a given frame with respect to the laboratory frame, $v_x$
its projection on the $x$ axis. The Lorentz transform allows us to
express $v_{QI,min}$ as a function of the values measured in the
laboratory frame: \ba \left.
\begin{array}{ccc} x&=&\gamma(x_{Lab}-v_xt_{Lab})\\
t&=&\gamma(t_{Lab}-v_xx_{Lab}/c^2)
\end{array} \right\}&\rightarrow& v_{QI,min}(v_x)\,=\,
-\frac{d_{AB}+v_x\tau}{\tau+v_xd_{AB}/c^2}\ea where
$\gamma=(1-v^2/c^2)^{-1/2}$, $d_{AB}=x_{B,Lab}-x_{A,Lab}$, which is positive by our convention,
and $\tau=t_{A,Lab}-t_{B,Lab}$, which is positive if the
detection at $A$ occurs, in the laboratory frame, after the
detection in $B$. Introducing the dimensionless parameters
$\beta_x=v_x/c$, describing the boost, and
$r=c\tau/d_{AB}=c/|v_{QI,min}(Lab)|$, describing the setup in the
laboratory frame, we find the expression \be
v_{QI,min}(\beta)\,=\, -c\, \frac{1+r\beta_x}{r+\beta_x}\,.
\label{vmin}\ee Thus $v_{QI,min}(\beta)$ depends on the orientation
of the setup through $\beta_x$, and on the
{\em relative} precision of the alignement $r$. It can be checked
that $|v_{QI,min}(\beta)|>c$ if and only if $|v_{QI,min}(Lab)|>c$
($r<1$), independent on $\beta$; this is indeed a necessary
property of the speed of quantum information, since two events
that are space-like separated cannot be found to be time-like
separated in any frame. From now on, we suppose that we consider
space-like separated events ($r<1$).

We conclude this section by noting that, due to finite accuracy in
the experiments, one can never conclude to perfect simultaneity.
In an EPR experiment, the localization $\Delta\tau$ of the photons is a limit
to the accuracy; the best localization that can be achieved is the
coherence length $\tau_c$ . The upper limit of
$|v_{QI,min}(\beta)|$ that we can reach is thus $d_{AB}/\tau_c$ in
all frames. The next section introduces a particular moving frame,
the CMB-frame.

\section{Linking the G-frame to the CMB-frame.}

As emphasized explicitely in several works, even recently
\cite{caban,mau,har}, the frame of the cosmic microwave background
radiation (CMB-frame) is a natural candidate for playing the role
of PF. The CMB-frame is defined as the frame in which the cosmic
background radiation is isotropic \cite{url}.

\begin{figure}
\begin{center}
\epsfxsize=15cm \epsfbox{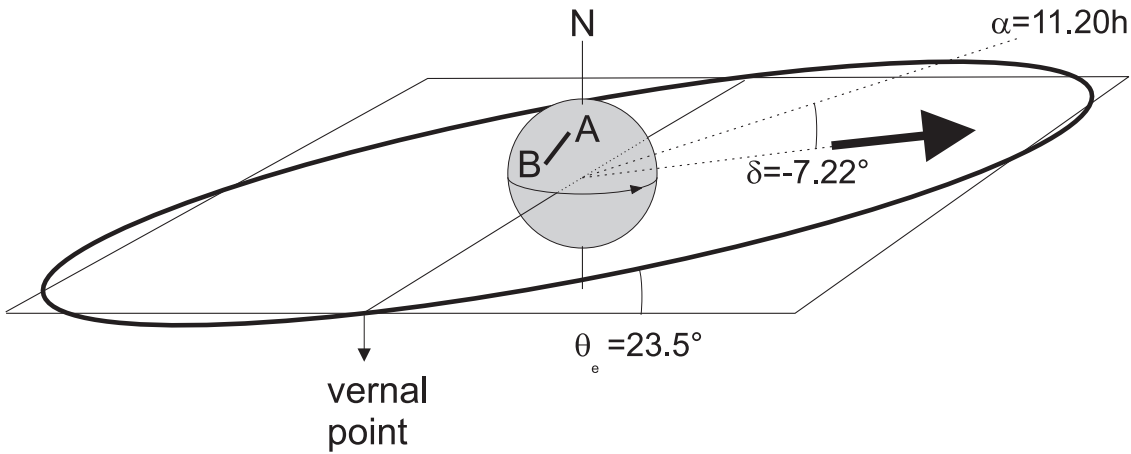}
\caption{Schematic
representation of the direction of $\vec{v}_{S)CMB}$ (black
arrow). The sphere is the Earth, A is Bellevue and B is Bernex.
The plane (celestial equator) contains the equator of the Earth;
the black curve is the trajectory of the Sun as seen from the
Earth (eclyptic). The vernal point is defined as the intersection
of the celestial equator and the eclyptic at the Spring equinoxe.}
\label{figastro}
\end{center}
\end{figure}

To find the relative speed of the G-frame with respect to the
CMB-frame, we need to take into account: (i) The speed $\vec{v}_{S)CMB}$ of the
barycenter of the solar system (identified with the Sun for our
purposes) with respect to the CMB; (ii) the speed of the Earth
with respect to the Sun $\vec{v}_{E)S}$, and (iii) the spin of the
Earth, giving the speed of Geneva with respect to the center of
mass of the Earth $\vec{v}_{G)E}$. The speed $\vec{v}_{S)CMB}$ is given in the literature
\cite{cmb}: its magnitude is ${v}_{S)CMB}=371$ km/s; its
direction in the orthogonal celestial coordinates (see fig. \ref{figastro}) is
$(\alpha=11.20^{h},\delta=-7.22^{\circ})$. For the other two speeds, we have in magnitude:
$v_{E)S}=2\pi D_{\oplus}/(1\,year)\approx
30$ km/s, where $D_{\oplus}$ is the distance Earth-Sun; and $v_{G)E}= 2\pi
R_{\oplus}\cos l_G/(24\,hours)\approx 0.31$ km/s, where
$R_{\oplus}$ is the Earth's radius and $l_G\approx 43^{\circ}$ is
the latitude of Geneva. The three speeds being much smaller than
$c$, we can use the Galilean addition rule: \be
\vec{v}_{G)CMB}\,=\,
\vec{v}_{G)E}+\vec{v}_{E)S}+\vec{v}_{S)CMB}\,. \ee Due to its
magnitude, the correction $\vec{v}_{G)E}$ can be neglected in
numerical estimates. Of course, the speed of the CMB-frame with
respect to the G-frame is given by
$\vec{v}_{CMB)G}=-\vec{v}_{G)CMB}$.

To go further, we need to introduce a system of coordinates, which
can be chosen arbitarily. We choose a cartesian coordinate system
defined as follows: the $z$-axis is the North-South axis of the
Earth, oriented in the N direction. The celestial equatorial plane
is therefore the $(x,y)$ plane. In this plane, the $x$ direction
is chosen to be the direction of the vernal point, that is the
point where the eclyptic intersects the celestial equatorial plane
at the Spring equinoxe (see fig. \ref{figastro}). Thus: \be
\vec{v}_{S)CMB}\,=\, {v}_{S)CMB}\,\left(
\begin{array}{c}\cos\phi_v\,\sin\theta_v\\ \sin\phi_v\,
\sin\theta_v\\ \cos\theta_v \end{array}\right)\ee with
$\phi_v=11.20h=168^{\circ}$ and $\theta_v=97.22^{\circ}$
\cite{cmb}. Also, neglecting the excentricity of the Earth's
orbit, it is not difficult to show that \be
\vec{v}_{E)S}\,=\,\omega_y\,D_{\oplus}\,\left(
\begin{array}{c}-\sin(\omega_y t+\theta_0)\\
\cos(\omega_y t+\theta_0)\,\cos\theta_e\\
-\cos(\omega_y t+\theta_0)\,\sin\theta_e \end{array}\right)
\label{ves}\ee where $\omega_y=\frac{2\pi}{1\,year}$ and
$\theta_e=23,5^{\circ}$ is the inclination of the eclyptic plane
with respect to the equatorial plane.

We still have to define the origin of time. It is natural to set
$t=0$ at the beginning of the EPR experiment we want to analyse.
Since the acquisition time is typically some hours, $\omega_y t$
in (\ref{ves}) can be set to 0. The definition of $t=0$ provides
also the interpretation of $\theta_0$: this angle is defined as
$\omega_y \Delta T$, with $\Delta T$ the time elapsed since the
Spring equinoxe at the moment of the experiment.

Recall that in eq. (\ref{vmin}) we must enter the projection
of $\vec{v}_{CMB)G}$ on the direction defined by the straight line
joining the two detectors. In our coordinate system, this
direction is given by \be
\hat{e}_x\,\equiv\,\frac{\vec{AB}}{|\vec{AB}|}\,=\,\frac{1}{N}\,\left(
\begin{array}{c}\sin\theta_B\cos\phi_B\,-\,\sin\theta_A\cos\phi_A\\
\sin\theta_B\sin\phi_B\,-\,\sin\theta_A\sin\phi_A\\
\cos\theta_B\,-\,\cos\theta_A
\end{array}\right)\,. \label{veab}\ee where $N=\sqrt{2}[1-\cos\theta_A\cos\theta_B-
\sin\theta_A\sin\theta_B\cos(\phi_A-\phi_B)]^{1/2}$. If $A$ is
Bellevue ($46^\circ 15'$N, $6^\circ 09'$E) and $B$ is Bernex
($46^\circ 10'$N, $6^\circ 05'$E), then we have:
$\theta_A=43^\circ 45'$, $\theta_B=43^\circ 50'$,
$\phi_A=\phi_0+\omega_d t$, $\phi_B=\phi_0+\omega_d t-0^\circ
04'$. The angle $\phi_0$ measures the position of the vernal point
with respect to the meridian of Bellevue at the beginning of the
experiment.

\section{Analysis of experimental data}

For the study of the speed of quantum information with an EPR
setup, the precision of the alignment $|r|=c\tau/d_{AB}$ is the
figure of merit. In fact, looking at eq. (\ref{vmin}), we see that the
simultaneity condition $r+\beta_x=0$ can be satisfied only if the
precision of the alignment $r$ satisfies $|r|<\mbox{max}_t
|\beta_x(t)|\approx |\vec{v}|/c$. Thus, the smaller the speed of
the considered frame with respect to the laboratory frame, the
higher the precision required to satisfy the simultaneity
condition. In other words, for a given frame, two situations may
arise: (i) The situation of {\em bad alignment} is described by
$|r|>\mbox{max}|\beta_x|$. In this case,
$|v_{QI,min}(\beta)|\approx\frac{c}{|r|}$. (ii) The situation of
{\em good alignment} is the opposite one: the simultaneity
condition can be satisfied. $|v_{QI,min}(CMB)|$ is no more limited
by $r$, but there are still two possible limiting factors. The
first one is the localization of the photons, that we have already
mentioned. To discuss the second one, we begin by noting that
$\beta_x$ varies with time due to the rotation of the Earth around
its axis, since the line Bellevue-Bernex is not parallel to this
axis --- and in fact, $\hat{e}_x$ given by eq. (\ref{veab}) varies
with $t$ through $\phi_A$ and $\phi_B$. In particular, $\beta_x$
is not constant during the time needed to record an interference
fringe. The speed $v_{QI,min}(\beta)$ depends on $\beta_x$ and on
$r$, so in principle one could keep it constant even though
$\beta_x$ changes, "simply" by performing the suitable continuous
correction of the alignment $r$. In practice, such a protocol is
cumbersome. So $v_{QI,min}(\beta)$ also varies during the
recording of a fringe. This second limiting factor is actually the
most important one, as will be shortly shown.

If the considered frame is the CMB-frame, whose motion is rather
slow since $|\vec{v}_{CMB)G}|\approx 300$ km/s, then a good
alignment is obtained for $|r|\sim 10^{-3}$. Since such a
precision was not looked for, the alignment was probably "bad" in
most of the previously reported EPR experiments \cite{note4}. In
the experiment that we are going to consider \cite{roue}, the
distance between the detectors was 10.6 km, and the difference in
the two arms was lowered down to 1-10 mm, whence $|r|\geq
10^{-6}$: the "good alignment" criterion for the CMB-frame is
clearly fulfilled. Due to chromatic dispersion in the fibers, the
localization of the photons was
$\Delta\tau=90$ ps, so that the maximal value of $|v_{QI,min}(CMB)|$
that we can hope to obtain is about $3.5\times 10^5\,c$
\cite{note2}. It took about one hour to record a fringe, the
detection rates being lowered by the photon losses in the fibers.

\begin{figure}
\begin{center}
\epsfxsize=15cm \epsfbox{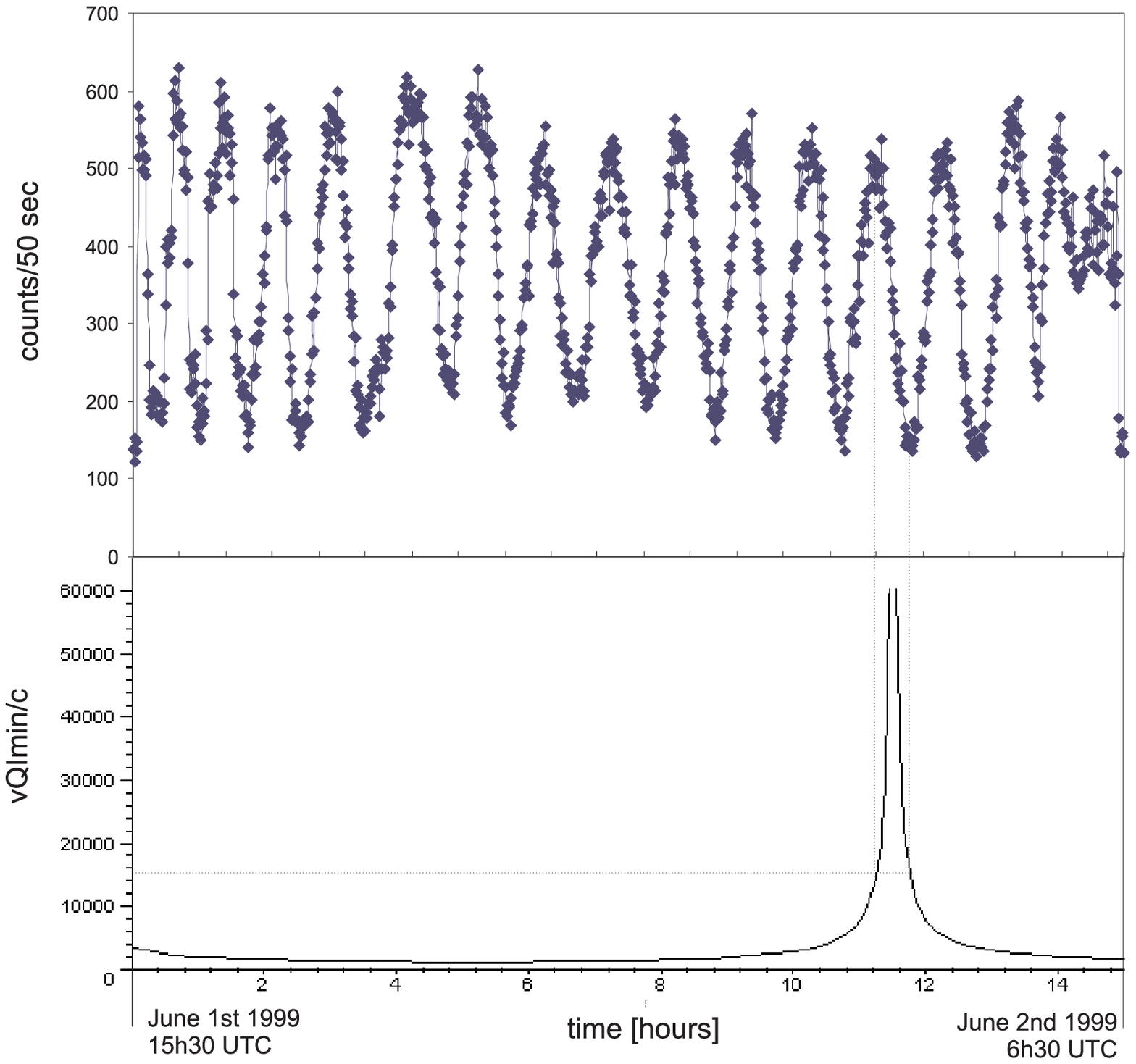}
\caption{(a) Two-photon interference fringes measured over 15
hours, each data point corresponds to a time interval of 50 s. (b)
The value of $|V_{QI,min}(CMB)|$ calculated by eq. (\ref{vmin})
for the day and hours of the experiment.} \label{figtrace}
\end{center}
\end{figure}

A typical EPR correlation trace is given in fig. \ref{figtrace}
(a). This data acquisition started on June 1st 1999 at 15h30 UTC
(whence $\theta_0=1.24$ rad and $\phi_0=2.247$ rad), and ended on
June 2nd at 6h30 UTC. Slight variations of the temperature led to
a variation of the length of the fibers linking the source to the
analyzing stations, that is, to a variation of the alignment
$\tau$ in the G-frame. We did not monitor $\tau(t)$ continuously;
we assume a linear interpolation between the
initial value $c\tau_i=2$ mm, and the final value $c\tau_f=14$ mm.
This assumption completes the set of numerical values needed to
evaluate $|v_{QI,min}(CMB)|$ according to formula (\ref{vmin}), as
a function of time, that is, as a function the Earth's rotation
around its axis. The numerical evaluation is shown in fig.
\ref{figtrace} (b). We see that at a given moment (about 3h UTC)
the detection events were simultaneous in the CMB-frame. Even at
that moment, the visibility of the fringes is not reduced. By
requiring the reduction of a whole half-fringe as a conservative
criterion for fixing a limit to the speed of quantum
information, we find the lower bound $|v_{QI,min}(CMB)|=1.5\times
10^4\,c$.

\section{Conclusions}

We have presented the first analysis of the results of an EPR
experiment in the frame of the cosmic microwave background
radiation. The conservative bound that we obtained for the "speed
of quantum information" in that frame, $|v_{QI,min}(CMB)|=
1.5\times 10^4\,c$, is still quite impressive, but, like most
physicists, the present authors will not be astonished if further
experiments provide an even higher value. The method of analysis
that has been developed in this work could of course be applied
to all possible frames \cite{note5}.

In the experiment that we analyzed, the recording of the fringes
is slow compared to the variation of $v_x$ induced by the
rotation of the Earth, and this is the constraint that fixes the
above limit of $|v_{QI,min}(CMB)|$. The precision required for
some planned experiments with laboratory distances \cite{aom}
should increase the bound $|v_{QI,min}(CMB)|$ up to $5\times
10^5\,c$, the rotation of the Earth still being the most
important limiting factor. As an order of magnitude, we estimated
that a fringe should be completed in less than 5 seconds in order
to reach the limit imposed by the localization of the photons.

We acknowledge discussion with R. Durrer, I. Percival and Ph.
Eberhard. We acknowledge financial support of the "Fondation Odier
de psycho-physique" and the Swiss National Science Foundation.

\newpage

\begin{thebibliography}{10}
\bibitem{note0} Obviously, we don't discuss here historical
versions of the PF (the "ether" of the XIX century), for which
experiments have been designed and performed. The requirement,
that a PF for quantum phenomena must not predict wrong results for
classical experiments, can be fulfilled \cite{reuse,caban}.
\bibitem{note1} We call "speed of quantum information" what is
probably better known as "speed of the
spooky-action-at-a-distance", in Einstein's terms.
\bibitem{note2} In some experiments described in \cite{roue}, a filter was used to
reduce the effect of the chromatic dispersion, leading to $\Delta\tau=5$ ps. We thus obtained
in the G-frame $|v_{QI,min}(Lab)|\approx\frac{2}{3} \,10^7\,c$.
For our present analysis we don't need such a precision, since
the spin of the Earth is the most important limiting factor.
\end{thebibliography}
\end{document}